# Metasurface Holographic Optical Traps for Ultracold Atoms

Xiaoyan Huang[1†], Weijun Yuan[2†], Aaron Holman[2], Minho Kwon[2], Stuart J. Masson[2], Ricardo Gutierrez-Jauregui[2], Ana Asenjo-Garcia[2], Sebastian Will[2*], and Nanfang Yu[1*]

1. Department of Applied Physics and Applied Mathematics, Columbia University, New York, New York 10027, USA
2. Department of Physics, Columbia University, New York, New York 10027, USA

† denotes equal contributions, * denotes corresponding authors

**Abstract**: We propose metasurface holograms as a novel platform to generate optical trap arrays for cold atoms with high fidelity, efficiency, and thermal stability. We developed design and fabrication methodologies to create dielectric, phase-only metasurface holograms based on titanium dioxide. We experimentally demonstrated optical trap arrays of various geometries, including periodic and aperiodic configurations with dimensions ranging from 1D to 3D and the number of trap sites up to a few hundred. We characterized the performance of the holographic metasurfaces in terms of the positioning accuracy, size and intensity uniformity of the generated traps, and power handling capability of the dielectric metasurfaces. Our proposed platform has great potential for enabling fundamental studies of quantum many-body physics, and quantum simulation and computation tasks. The compact form factor, passive nature, good power handling capability, and scalability of generating high-quality, large-scale arrays also make the metasurface platform uniquely suitable for realizing field-deployable devices and systems based on cold atoms.

## 1. Introduction

Single atoms in optical trap arrays provide a promising platform to conduct fundamental quantum optics experiments and can enable a variety of technical applications, such as quantum metrology, quantum simulation, and quantum computation. Conventional approaches to generate trap arrays have relied on acoustic optical diffractors (AODs) [1,2], liquid crystal-based spatial light modulators (SLMs) [3-5], and digital micromirror devices (DMDs) [6]. However, these approaches are associated with a number of drawbacks, including large device footprints, stringent requirements on power supply and cooling, lack of control of the polarization state of light, and limitations in the size and geometry of generated trap arrays. Alternatively, a few passive devices, such as amplitude masks [7] and microlens arrays [8], have been developed and demonstrated to successfully trap arrays of atoms. These methods still require bulky optical components for demagnifying and projecting the trap array patterns. Next-generation quantum devices based on atomic arrays require a compact and scalable solution that is able to generate arbitrary trapping geometries with a minimum of free-space optical components.

Here, we propose dielectric metasurface holograms as a new platform to efficiently and faithfully generate optical traps with desired geometries (**Fig. 1**). Metasurfaces are composed of a 2D array of meta-units and offer complete, independent, and precise manipulation of optical amplitude, phase, and polarization across the wavefront with subwavelength resolution [9-11]. Recent efforts in incorporating metasurfaces into cold atom setups have led to the demonstration of a metasurface optical chip that can generate a 3D magneto-optical trap (MOT) for Rubidium atoms with a single incident laser beam [12]. In another demonstration, a metasurface lens has been used to focus down a 3×3 array of laser beams generated by an AOD and Rubidium atoms have been subsequently trapped into the array [13]. However, to our knowledge, there has been no report of metasurfaces that combine generation and focusing of trap arrays into a single device. In this work, we explore the capability of metasurface holograms in directly producing optical trap arrays for single atoms, including complex trap geometries such as quasi-crystals, kagome lattices, and twisted bilayers. Our passive metasurface devices are shown in experiments to exhibit high efficiency, high accuracy, and superior power-handling capabilities.

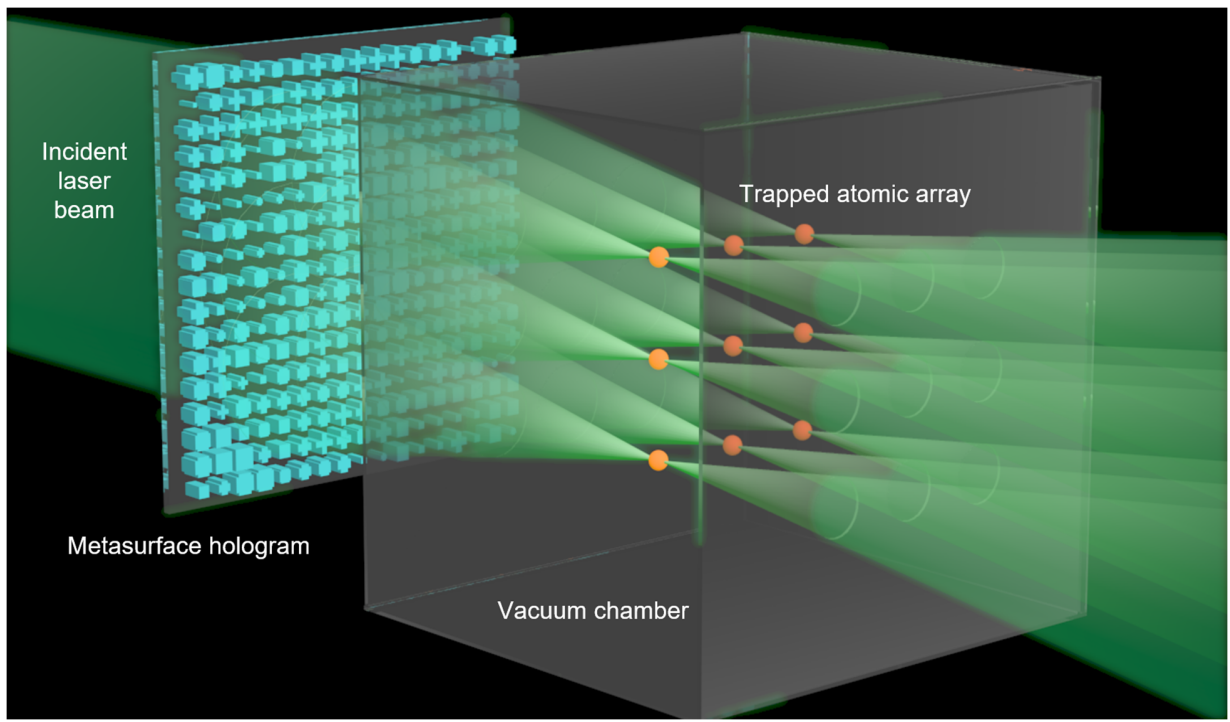

**Figure1**: Schematic illustration of a compact system to generate optical trap arrays for cold atoms using a metasurface hologram. The metasurface modulates the incident laser beam without the assistance of any other optical components and forms an optical trap array in the vacuum chamber. A 3×3 square lattice array is illustrated but the geometry of the trap array can be arbitrary.

The structure of the remainder of the paper is as follows. **Section 2** outlines fundamental problems that can be investigated using atomic arrays, as well as potential applications. **Section 3** briefly reviews the physics and recent development of metasurfaces. **Section 4** discusses strategies we developed to design and fabricate phase-only meta-holograms for atomic array experiments. **Section 5** presents the generated optical trap arrays and their characterization. **Section 6** discusses future prospects of integrating holographic metasurfaces into atomic array experimental setups. In **Section 7** we draw our conclusions.

## 2. Single atoms in optical trap arrays: a scalable platform for quantum science and applications

A long-standing challenge in quantum physics is to prepare large systems of individually addressable qubits. One promising platform consists of arrays of trapped single atoms. Early demonstrations of this idea involved creating an atomic Mott insulator using an optical lattice potential [14-17], where the condition of a single atom per site is enforced by strong on-site repulsive interactions. More recently, optical trap arrays have been used to create arrays of neutral alkali atoms [18-20], Strontium [21,22] and Ytterbium [23] atoms, and even molecules [24].

Atomic arrays provide a versatile interface for controlling light-matter interactions. Photon-mediated interactions have been experimentally demonstrated in 1D and 2D atomic arrays [25,26]. The optical properties of the array, as a result of these interactions, are collective, leading to exotic quantum mechanical properties fundamentally different from those of independent atoms. For example, in ordered structures with a small lattice constant, destructive interference leads to sub-radiant states [27,28]. These states are extremely long-lived and can be used for improved quantum memories [29] and lossless transport of light [30,31]. Atomic arrays have been suggested as light sources with unconventional properties, such as anti-bunched light [32,33], directional single photons [34,35], and extremely bunched light [36,37]. In the last couple of years, a number of novel applications have emerged, the most noticeable of which has been atomically thin mirrors that are switchable by controlling just a single atom [38]. Moreover, for systems with a high density of excitations, their decay becomes correlated at the many-body level. For instance, when atoms are fully excited, they start synchronizing as they decay, which leads to super-radiant emission in the form of a short bright pulse of light [39,40]. At later times, the atoms become correlated due to the population of long-lived sub-radiant states [41-43].

Atomic arrays have great potential in metrological applications [44,45]. Optical clocks with incredible precision and accuracy have been achieved by single ions and neutral atoms trapped in optical lattices. The former allows for addressing individual ions, which permits high duty-cycle interrogation. The latter achieves low atom shot noise by interrogating thousands of atoms at the same time. An atomic array, observed via single-atom resolved detection, combines the merits of these platforms as it allows one to correct for systematic effects on the single-atom level, while having the potential to scale up to relatively large array sizes. In addition, atomic arrays suppress collisional shifts and tunneling-induced shifts that are prevalent in more traditional optical lattice clocks, and they allow the implementation of entanglement schemes that have the potential to achieve clock-precision beyond the standard quantum limit. Very recently, it has been shown that an atomic clock based on a 2D array of Strontium atoms can achieve a coherence time of 48 seconds, substantially exceeding the previous state of the art [46]. In addition to improving the metrological precision and accuracy of optical clocks, intense efforts are currently under way to build portable, compact setups to suit real-life application scenarios [47-51]. Here, metasurface generated arrays that are mechanically stable and do not consume power promise to play a critical role towards field-deployable, compact atomic clocks.

Atomic arrays can also be used as quantum simulators to implement Hamiltonians of a plethora of spin models. Interactions between atoms can be engineered via Rydberg excitations [52], where electrons in high principal quantum number orbitals interact via van der Waals forces. In particular, the effect of Rydberg blockade, where a Rydberg excitation on one atom blocks excitations on its neighbors, has been used to simulate Ising-type spin models [5,53-56], the Su-Schrieffer-Heeger model [25], and frustrated magnetism [57]. Other noteworthy developments include the demonstration of topological states [25,57], quantum many-body scars [54], and time crystal-like behavior [58]. With arbitrary control of the trap array geometry comes the ability to build interacting atomic systems with exotic lattice structures inspired by other disciplines of research, such as condensed matter and material physics [59,60], high-energy physics and cosmology models [61-65], and quantum chemistry [66], making atomic arrays a powerful quantum simulation platform of universal interest to the broad scientific community.

Moreover, arrays of single neutral atoms constitute a promising avenue for quantum computation, with each atom playing the role of a single qubit. Quantum gates can be realized by using control fields and Rydberg states [67,68]. Site-resolved optical addressing produces an AC Stark shift of the resonance frequency of target atoms in an array such that they become resonant with a global field that mediates the interactions between them. Single-qubit gates [69,70] and controlled-phase gates for two qubits [71,72] have been demonstrated based on such principles. Great efforts are being spent in both academia and industry to explore atomic arrays as building blocks of a universal quantum computer [73].

As illustrated by the above examples, the arrangement of atoms and the nature of their interactions are determined by the configuration of the optical traps (**Table 1**). In forming atomic arrays, it is essential to generate optical trapping potentials with desired geometric configurations, while maintaining high optical efficiency, positioning accuracy, and intensity uniformity of the traps.

| Application examples | Geometries | Interatomic spacing |
|---|---|---|
| Quantum magnetism | Kagome, triangular, honeycomb lattice etc. | A few microns |
| Quantum optics | Various 1D, 2D, and 3D geometries | Smaller than or comparable to excitation wavelength |
| Atomic clock | Square lattice etc. | As small as 1 micron |
| Quantum computing | Dimerized traps, square lattice, 3D cubes | A few microns |

**Table 1**: Applications of neutral atom arrays and the relevant geometries and spacing for each application.

## 3. Metasurfaces: Arbitrary wavefront shaping with micron-thick, nanostructured thin films

In recent years, metasurfaces have emerged as a powerful platform to shape and manipulate optical waves [74-79]. Metasurfaces are composed of 2D arrays of optical scatterers (dubbed "meta-units"), the sizes and shapes of which can be tailored to control amplitude, phase, and polarization over the optical wavefront. Metasurfaces have a planar form factor: the height of meta-units, typically ranging from a few hundred nanometers to one micron, is small compared to the linear dimension of a metasurface, which ranges from tens of microns to several centimeters [80,81]. The subwavelength cross-sectional sizes of meta-units allow for high forward-scattering efficiency and molding the optical wavefront with high spatial resolution.

The control of optical phase by dielectric metasurfaces is typically realized through dispersion engineering or polarization conversion [10,11,82-89]. In the former case, meta-units are treated as short waveguide segments standing on a substrate. They are designed to possess various cross-sectional shapes, and thus support waveguide modes with different modal indices, depending on the spatial overlap between the mode and the dielectric material. This results in controllable phase accumulation as the mode propagates through the meta-units [10,86,88]. In the latter case, the meta-units are designed to be optically birefringent (i.e., with anisotropic cross-sections): circularly polarized light is converted into the opposite handedness and picks up a phase proportional to twice of the orientation angle of the anisotropic meta-units, an effect known as geometric or Pancharatnam-Berry phase [11,88]. Both methods are capable of delivering phase modulation over the entire $2\pi$ range, and by combining the two design strategies, complete and independent control of two optical parameters (e.g., phase and amplitude) over the optical wavefront can be achieved. In the visible and near-infrared spectral ranges, metasurfaces made of dielectric materials can be highly transparent and exhibit robust performance under high power illumination [90].

The capability of controlling multiple optical parameters simultaneously with subwavelength resolution and the planar form factor of metasurfaces have enabled the creation of a variety of flat optical devices. Early experimental demonstrations of metasurface-based optical components include meta-lenses capable of producing diffraction-limited focal spots [91], phase plates that could generate optical vortex beams [77], and flat holograms [92], where the phase profile dictated by a desired holographic object was realized by metasurfaces. Recent work on metasurfaces has focused on demonstrating functionalities that cannot be easily realized by conventional refractive optics. For example, polychromatic and achromatic single-element meta-lenses have been demonstrated in both the visible [88,89] and near-infrared [86] spectral ranges by using dispersion-engineered meta-units with complex cross-sectional shapes. Monochromatic aberration correction has been achieved by cascaded meta-lenses [93,94]. Spatial multiplexing, polarization multiplexing, and resonant-mode multiplexing have enabled multiple distinct wavefronts to be encoded into a single metasurface, allowing one device to achieve multiple functionalities at once [10,11], or in a wavelength-or polarization-specific manner [9,95,96]. More recently, active metasurfaces have been extensively studied: by utilizing thermo-optical [97] and electro-optical [98] effects, and by incorporating MEMS [99] and phase-changing materials [100] into metasurfaces, reconfigurable flat optical devices, including varifocal lenses [101], parfocal zoom lenses [102], and amplitude modulators [103], have been demonstrated. Very recently, the capability of metasurfaces in conducting analog computing has been investigated and exciting progress has been made to create edge detectors [104] and optical neural networks that can classify simple objects [105].

In this work, we present the design and experimental demonstration of dielectric metasurface holograms to generate densely spaced optical traps with designer configurations from 1D to 3D and with high fidelity and efficiency. The metasurface hologram platform can potentially enable the study of exotic quantum optical phenomena hitherto not easily attainable with other platforms, and can substantially reduce the complexity, volume, and cost of atomic-array-based quantum devices, including atomic clocks, and quantum simulation and computation systems.

## 4. Design and Fabrication of Phase-only Meta-Holograms

We designed and realized both polarization-independent and polarization-multiplexed meta-holograms using $TiO_2$ nanopillars patterned on an optically thick fused silica substrate to operate at $\lambda$=520 nm, the "magical wavelength" for trapping Strontium atoms [106]. We started by obtaining the amplitude and phase responses of individual $TiO_2$ meta-units using rigorous coupled wave analysis (RCWA) calculations. The polarization-independent meta-holograms are composed of nanopillars that have cross-sectional shapes with 4-fold symmetry (i.e., squares and crosses) and provide fixed phase responses irrespective of the polarization state of light; the polarization-multiplexed meta-holograms consist of birefringent nanopillars, providing independent control of optical phase at the two orthogonal polarization states. The calculated phase responses for the two meta-unit libraries are shown in **Figs. 2(a)** and **(b)**, respectively. Because of the low absorption of $TiO_2$ in the visible spectrum, both libraries have an average optical transmission of more than 90%.

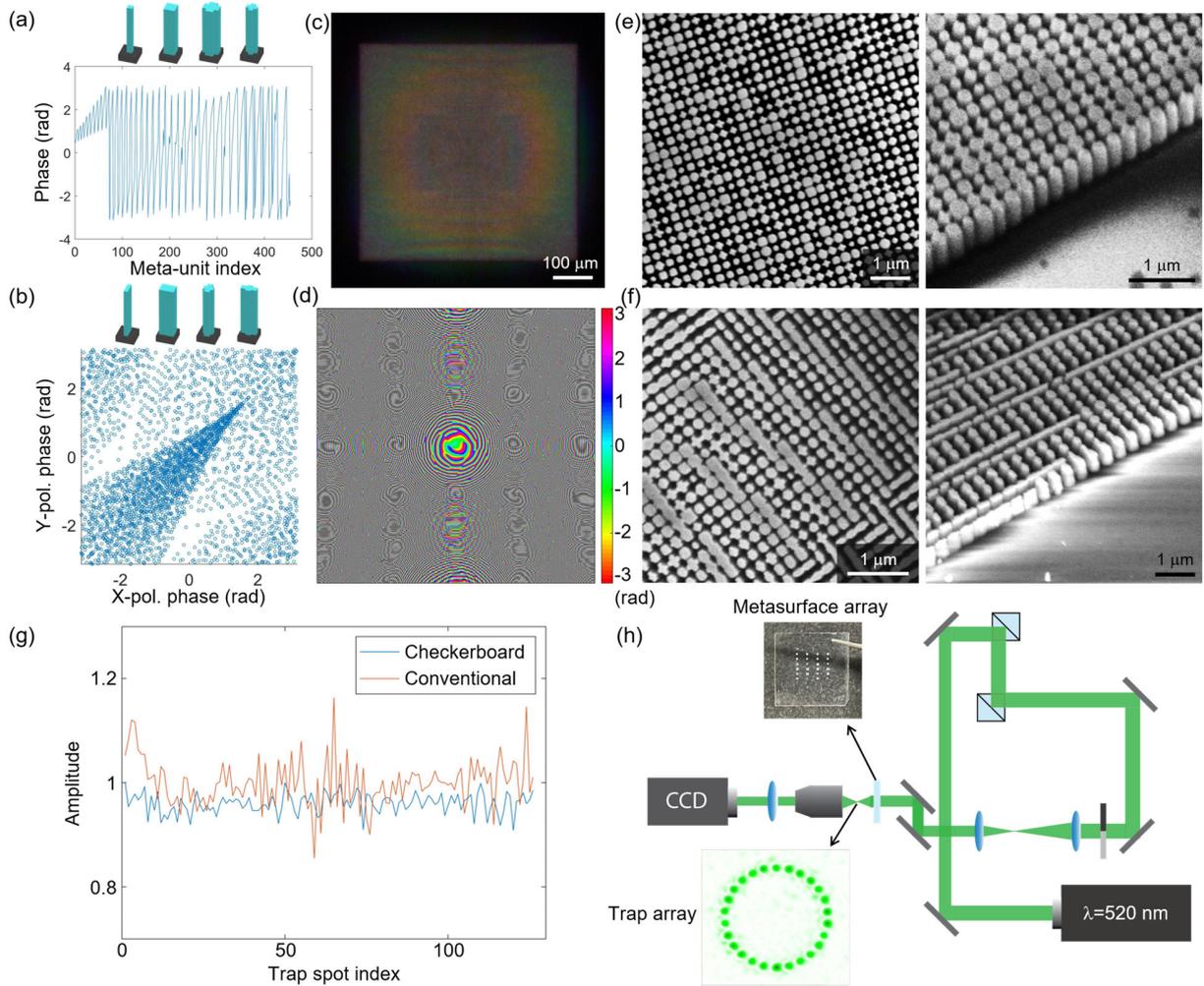

**Figure 2**: (a) Calculated phase responses of a library of 454 polarization-independent meta-units. The whole library can cover the $2\pi$ phase range multiple times. The meta-units have square and cross-shaped cross-sections, and are ordered and indexed by their dimensions. A few examples are shown on the top. (b) Calculated phase responses of a library of 3,126 birefringent meta-units. The two axes denote phase responses at two orthogonal polarization states, respectively. The library provides a dense sampling over the entire 2D phase space, demonstrating complete and independent control of phase at the two polarization states simultaneously. The meta-units have asymmetric cross-sectional shapes and a few examples are shown on the top. (c) Dark-field microscope image of a fabricated metasurface hologram for generating the 1D dimerized trap in **Fig. 3(a)**. (d) Calculated phase response for generating the ring array with spacing between adjacent spots of 1.5 $\mu$m in **Fig. 3(b)**. (e) SEM images of a fabricated polarization-independent metasurface. Left: top view; right: perspective view. (f) SEM images of a fabricated birefringent metasurface. Left: top view; right: perspective view. (g) Simulated

peak intensities of a ring array with 1.5-μm spacing using the checkerboard method (blue) and the conventional approach (red). (h) Schematic illustration of the experimental setup used for optical characterization of trap arrays.

An approach based on the Gerchberg-Saxton algorithm [107] was used to calculate the phase masks required for producing desired optical traps. In essence, the algorithm uses the Rayleigh-Sommerfeld diffraction integral to iteratively propagate back and forth between the metasurface plane and focal plane, while enforcing the phase-only condition upon the metasurface. In this way, it identifies the optimal phase profile that produces an intensity pattern on the focal plane that most closely matches the desired trap array (an example shown in **Fig. 2(d)**). We used a negative feedback routine in each iteration to further enhance the intensity uniformity of the traps [108]. Specifically, the target intensity of the (N+1)$^{th}$ iteration, $I_{N+1}(x,y)$, is determined by the simulated intensity of the $N^{th}$ iteration, $I_N(x,y)$, via

$$I_{N+1}(x,y) = \frac{I_{ideal}(x,y)}{I_N(x,y)},$$

where $I_{ideal}(x,y)$ is the ideal binary intensity distribution with unity amplitude at trap locations and 0 elsewhere. In this manner, traps brighter than intended (i.e., intensity larger than 1) are suppressed while traps darker than intended (i.e., intensity smaller than 1) are enhanced, resulting in a more uniform pattern.

By sweeping over all the lattice positions on the metasurface and choosing meta-units to minimize phase error locally, we can generate an optimal layout of meta-units with a collective phase profile that replicates the phase modulation prescribed by the Gerchberg-Saxton algorithm. For polarization multiplexed traps, this process was done in parallel for the two polarization states; that is, the choice of a meta-unit at a certain position of the metasurface in reference to the meta-unit library (**Fig. 2(b)**) has to minimize the local phase errors at both polarization states.

We employed a "checkerboard" technique for the generation of some arrays with closely spaced traps (e.g., 1D ring arrays, 2D square lattices): the entire array is split evenly into two parts in the spirit of a checkerboard dividing a square surface; the two parts are then realized by the two orthogonal polarization states, respectively. In this way, the difficult task of generating a large, closely spaced array is broken down to the generation of two sub-arrays, each with a smaller number of relatively loosely spaced traps. This strategy allowed us to realize large arrays with tight spacings between traps, while still maintaining high fidelity of individual traps and intensity uniformity among trap sites. In the example shown in **Fig. 2(g)**, we designed a ring array with 1.5-μm spacing using both the conventional method (without polarization multiplexing) and the checkerboard technique and compared the peak intensity variations of the simulated trap spots. The results show that the checkerboard technique significantly improved the quality of the array with a much weaker spot-to-spot intensity variation.

Metasurfaces designed using the above strategies were fabricated using standard, CMOS (complementary metal oxide semiconductor)-compatible nano-fabrication techniques. A thin film of $TiO_2$ of 800 nm in thickness was deposited by electron beam evaporation of $Ti_3O_5$ in an oxygen-rich atmosphere [109] on 500-μm thick fused quartz substrates. Electron beam lithography (Elionix ELS-G100) was conducted on a bilayer resist (PMMA 495k A4 and 950k A2) spun on top of the film with a dose of 770 μC/cm$^2$ at a current of 2 nA. A 20-nm layer of E-Spacer was spun on top of the double-layer resist to avoid the electron charging effect. The exposed resist was subsequently developed in an IPA:DI (3:1) solution for 2 minutes and coated with a bilayer etch mask of a 25-nm Cr film and a 15-nm $Al_2O_3$ film using electron beam evaporation. The mask was then lifted off in Remover PG overnight and the metasurface pattern was etched into the $TiO_2$ film in an inductively coupled plasma etcher (Oxford PlasmaPro 100 Cobra). Finally, the mask was removed after immersion in Cr etchant 1020 for 2 minutes. Our metasurface holograms have a linear dimension of ~400-560 μm and a numerical aperture (NA) of ~0.45, leading to a diffraction-limited trap spot size of ~500 nm at $\lambda$=520 nm. An optical image and scanning electron micrograph (SEM) images of fabricated metasurface holograms are shown in **Figs. 2(c)**, **2(e)**, and **2(f)**. The SEM images show that the nanopillars retain the designed cross-sectional geometries and have vertical sidewalls.

## 5. Results and Discussion

A schematic of the measurement setup is shown in **Fig. 2(h)**. The output from a $\lambda$=520 nm laser (Azurlight ALS-GR-520-5-A-CP-SF) was first treated with polarization optics and then modulated by a metasurface hologram. The generated optical trap array was imaged by a CCD camera equipped with an NA=0.6 objective. The effective magnification of the imaging system was calibrated by using the 1951 US Air Force resolution test chart, so that the geometry and spacing of generated trap spots can be accurately measured.

We experimentally realized a number of trap arrays (**Figs. 3-5**), thus demonstrating the ability of metasurfaces to generate arbitrary trapping configurations. Our demonstrated 1D arrays include a dimerized linear array with 5-μm spacing between two spots of the dimers and 10-μm spacing between adjacent dimers (**Fig. 3(a)**), and two ring arrays, one composed of 26 optical spots with 1.5-μm spacing between adjacent spots (**Fig. 3(b)**), and the other composed of 93 optical spots with 1.25-μm spacing between adjacent spots (**Fig. 3(c)**). Dimerized linear arrays have been previously utilized to achieve parallel control of multiqubit gates using Rubidium atoms and two-qubit gates using Strontium atoms with high fidelity [22,110]. For atoms arranged in a ring array, it has been predicted that spontaneous decay of excited atomic states can be exponentially suppressed, a promising platform for studying sub-radiant physics [28].

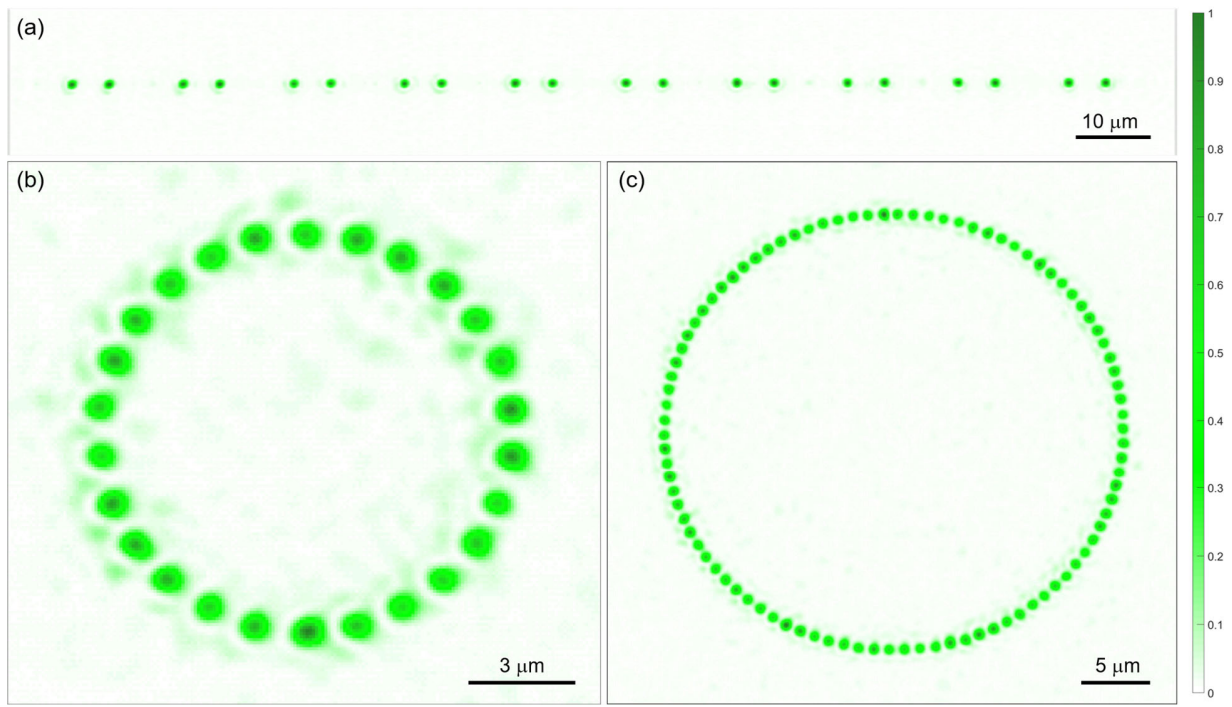

**Figure 3**: Measured intensity profiles of (a) a 1D dimerized trap array with 5-μm spacing between dimers, (b) a small ring array with 1.5-μm spacing, and (c) a larger ring array with 1.25-μm spacing. The intensity profile in each subfigure is normalized to its peak value.

Demonstrated 2D arrays include a 14×14 square array with a lattice constant of 1.8 μm (**Fig. 4(a)**), a kagome array composed of 300 optical spots with 5-μm spacing between the closest spots (**Fig. 4(b)**), and two Penrose-tiling type 2D quasi-crystal arrays, one composed of around 200 optical spots with the spacing between the closest spots of approximately 7.5 μm (**Fig. 4(c)**), and the other composed of double amount of optical spots (target spots and reservoir spots) with the spacing between the closest spots of approximately 4.5 μm (**Fig. 4(d)**). The reservoir spots will assist trapping cold atoms that can be subsequently moved to the target traps (as initial loading of the target traps by atoms is ~50%). Previously, kagome geometries with Rubidium atoms have been used to realize quantum spin liquid phases [57], and interesting magnetic orderings have been demonstrated in Penrose-tiling geometries [111,112].

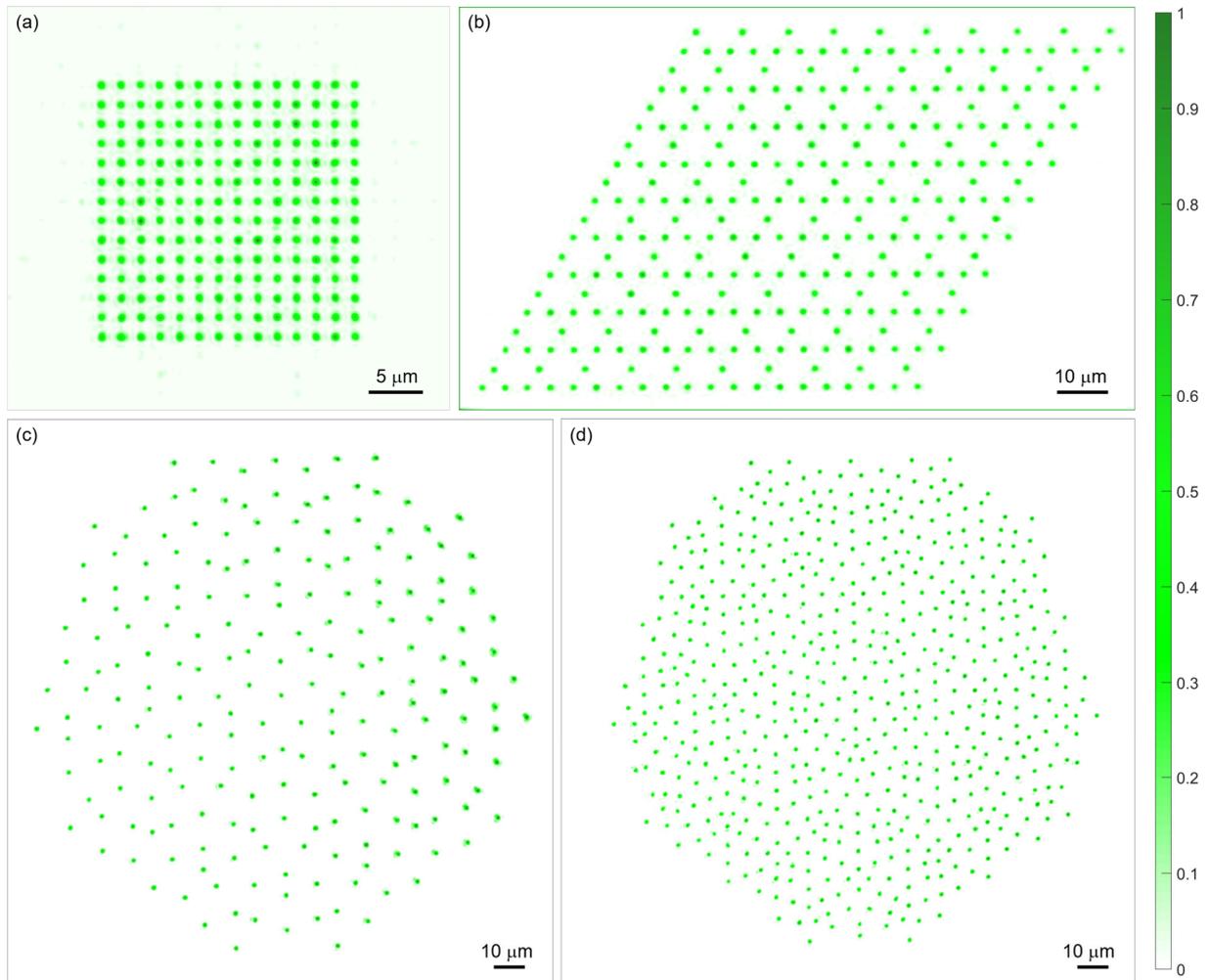

**Figure 4**: Measured intensity profiles of (a) a 2D square lattice with a lattice constant of 1.8 μm, (b) a kagome lattice with minimum spacing between adjacent spots of 5 μm, (c) a Penrose-tiling type quasi-crystal lattice with minimum spacing between adjacent spots of 7.5 μm, and (d) a Penrose-tiling type quasi-crystal lattice with reservoir traps and minimum spacing between adjacent spots of 4.5 μm. The intensity profile in each subfigure is normalized to its peak value.

Demonstrated 3D arrays include a cubic array composed of 147 optical spots (three layers, each containing 7×7 optical spots) with a lattice constant of 10 μm (**Fig. 5(d)**), and twisted bilayers consisting of either hexagonal or honeycomb 2D arrays (**Figs. 5(a)-(c)**), with 4-μm in-plane spacing between the closest spots and 10-μm inter-layer spacing. In the latter demonstrations, the twisting angle was 15 degrees for the hexagonal bilayer and 20 degrees for the honeycomb bilayer, and the two layers were generated, respectively, by the two polarization channels of a birefringent metasurface hologram. Moiré patterns were observed when the twisted bilayers were digitally overlapped on the same plot (**Figs. 5(b)** and **(c)**). It is expected that 3D cubic lattices can help enhance the connectivity between qubits in quantum computing devices based on neutral atoms [113]; twisted bilayer geometries with neutral atoms can provide a more controllable and flexible way to study twistronics compared with solid state systems [114,115].

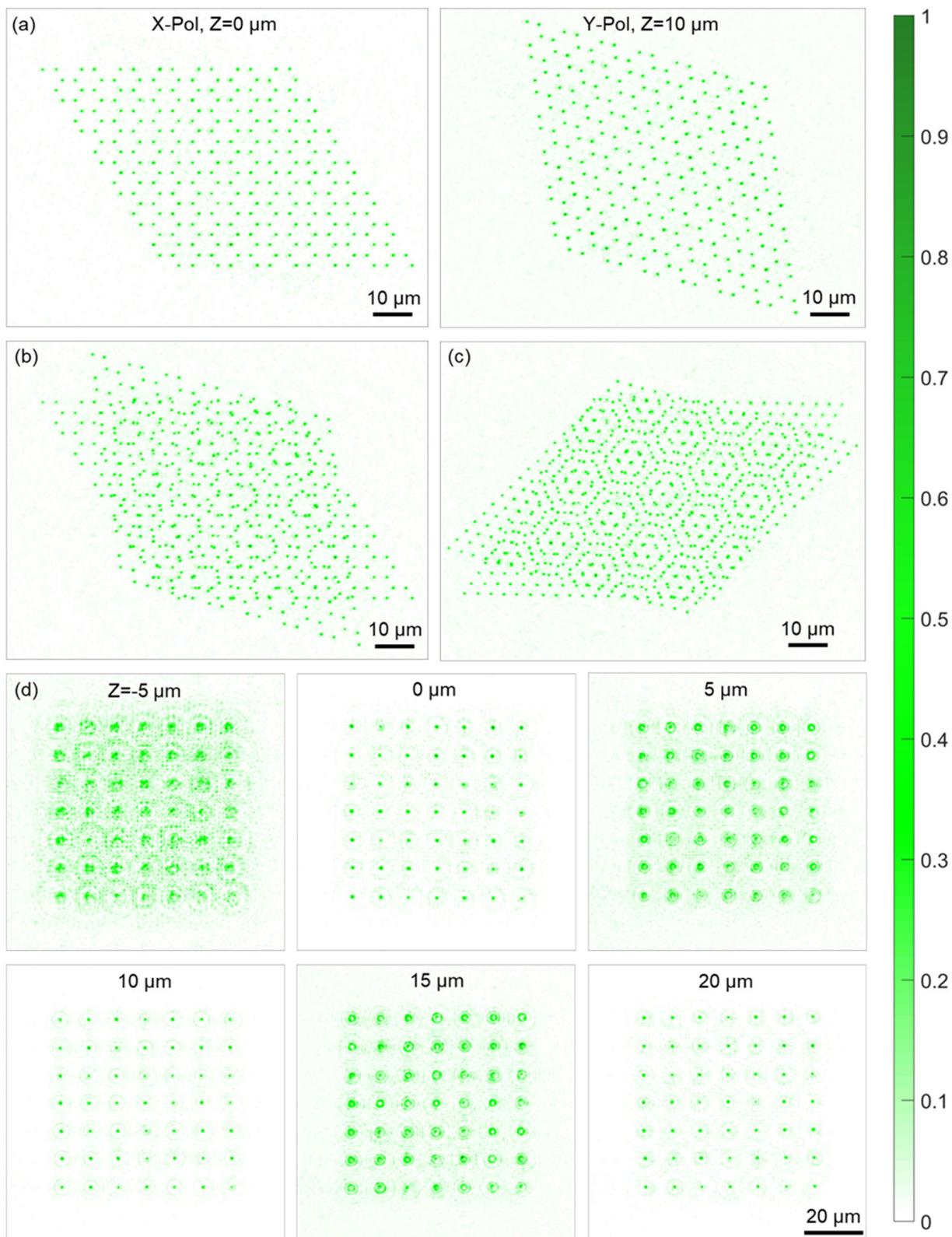

**Figure 5**: (a) Measured intensity profiles of a twisted bilayer consisting of two honeycomb lattices separated by 10 μm, which are encoded in two orthogonal polarization states of a metasurface hologram. (b) Digitally overlayed twisted honeycomb lattices, combining the images in (a) and showing a Moiré pattern. (c) Digitally overlayed twisted triangular lattices, generated by another metasurface hologram. (d) Measured intensity profiles of a 3D cubic lattice. Shown are far-field scans at different distances from a metasurface hologram generating the trap: three 7×7 square lattices are located at Z=0, 10, and 20 μm, respectively. The intensity profiles at different planes are normalized to their respective maxima; the maximum intensity at out-of-focus planes (Z=-5, 5, and 15 μm in (d)) is smaller than 5% of that at focal planes.

Optical efficiencies were measured for all the demonstrated trap arrays. Typical transmission efficiency, as defined by the ratio between power transmitted through a metasurface and power transmitted through a bare silicon dioxide substrate of the same size, is between 60%-70%, depending on the specific trap configuration. The focusing efficiency, defined as the fraction of power focused onto the intended trap sites versus that of the incidence, is generally between 40%-50%. In finite-difference time-domain (FDTD) simulations, the transmission efficiency is around 80% and the focusing efficiency 65-70%. The discrepancy between experiments and simulations likely originates from imperfect optical transparency of the $TiO_2$ films prepared, as well as errors in modeling and fabrication.

We can switch dynamically between two distinct array patterns produced by birefringent metasurface holograms by controlling the polarization of the input light. In the example shown in **Fig. 6(a)**, we demonstrated switching between an intact ring and a defective ring (with one optical spot displaced by a short distance out of the ring), when incident light was switched between two orthogonal linear polarizations. Atomic ring arrays are able to support deeply sub-radiant states, which can be accessed via local defects, because the defects facilitate the coupling with outside electromagnetic fields. The switching between perfect and defective rings can enable the realization of a quantum memory, where photons are stored and released via mechanical motion of a single atom.

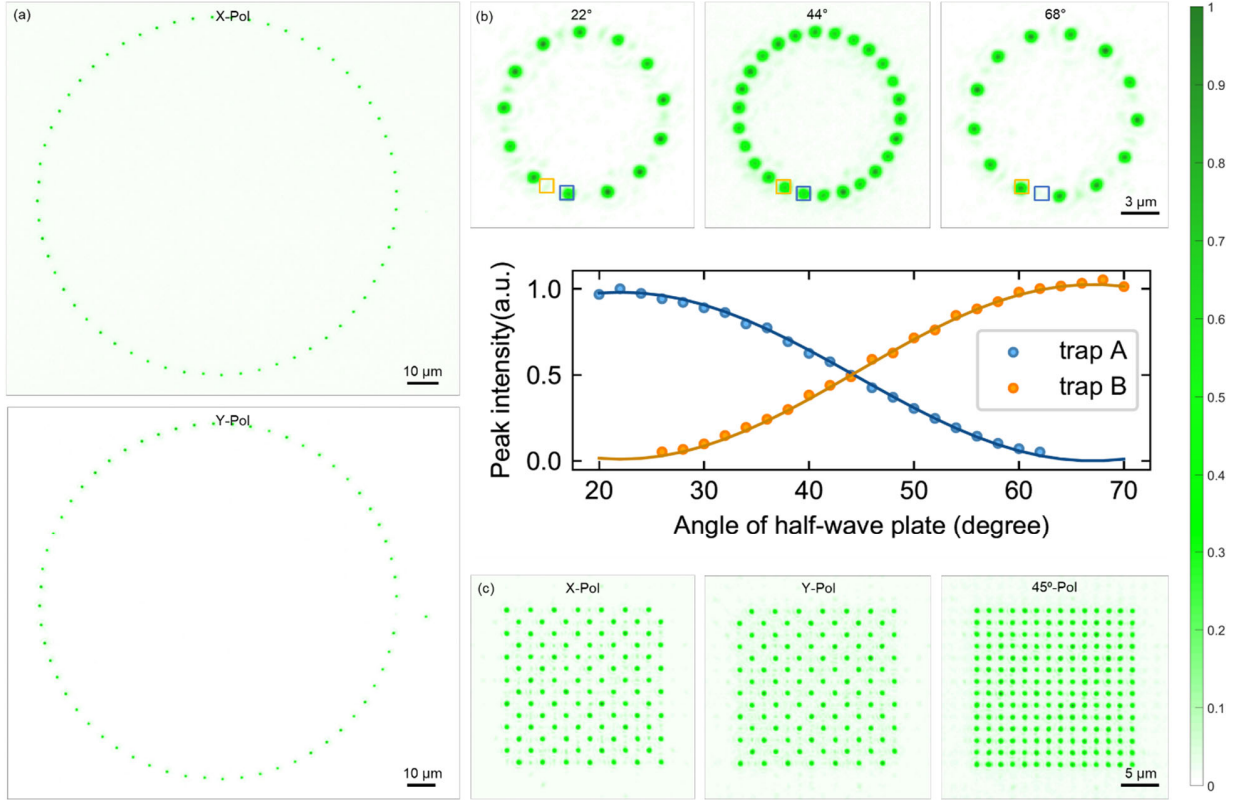

**Figure 6**: (a) Measured intensity profiles of a polarization-switchable ring array. The ring can be switched between an intact state (top) and a defective state (bottom) and can enable the realization of a quantum memory. (b) Evolution of a ring array as a function of the rotation angle of a half-wave plate, which controls the polarization angle of a linearly polarized incident laser beam. The top row shows the trap patterns at three specific polarization angles. (c) Measured intensity profiles of a polarization-multiplexed square lattice. Light at orthogonal polarization states generates spatially interleaved checkerboard patterns, which recombine into a square lattice at 45-degree incident polarization.

In a second example, we demonstrated a metasurface that can generate two ring arrays with an offset in the azimuthal angular direction at horizontal and vertical incident polarization states (**Fig. 6(b)**). Thus, when excited by incident light with 45-degree linear polarization, the metasurface produced a ring array with double the number of optical spots (i.e., the spacing between adjacent optical spots changes from 3 μm when excited by vertically or horizontally polarized light alone to 1.5 μm when excited by incident light with 45-degree linear polarization). This example is an implementation of the checkerboard technique mentioned in **Section 4**. Furthermore, continuously tuning the orientation of the incident linear polarization changed how the optical power was distributed between the two sub-rings. In **Fig. 6(b)**, we show the peak intensity of two adjacent optical spots in the ring array as a function of the rotation angle of a half-wave plate used to control the orientation of the incident linear polarization. The data is fitted to a function of $\cos^2(\theta)$, where $\theta$ is the rotation angle of the half-wave plate. **Figure 6(c)** shows a third example of a reconfigurable trap array where by tuning the incident polarization states, we were able to switch the array between a checkerboard pattern and a square pattern with a lattice constant as small as 1.8 µm.

To investigate the thermal stability of trap arrays generated by metasurface holograms, we conducted *in-situ* measurements of the arrays with extended periods of high-power illumination. For example, a metasurface that produces a ring array with spacing between adjacent spots of 2.5 µm was illuminated with a collimated CW laser beam at $\lambda$=520 nm with a beam diameter of ~300 µm and a power of 2.75 W continuously for 1.5 hours, and a CCD camera was used to monitor the generated trap array. The measured trap arrays at 0, 0.5, 1 and 1.5 hours into the experiment are presented in **Fig. 7**. We observed no degradation of the array throughout the testing period, a minor drift of the array along the vertical direction (<0.5 µm), and a drift along the horizontal direction of ~2.5 µm. The time constants obtained from exponential fits are 17 minutes for the vertical drift and 6.6 minutes for the horizontal drift. Besides thermal drifts, moderate defocusing of trap arrays was sometimes observed at the onset of the measurement. For example, the kagome array (**Fig. 4(b)**) was observed to drift along the light propagation direction by ~10 µm within the first 10 minutes of high-power illumination.

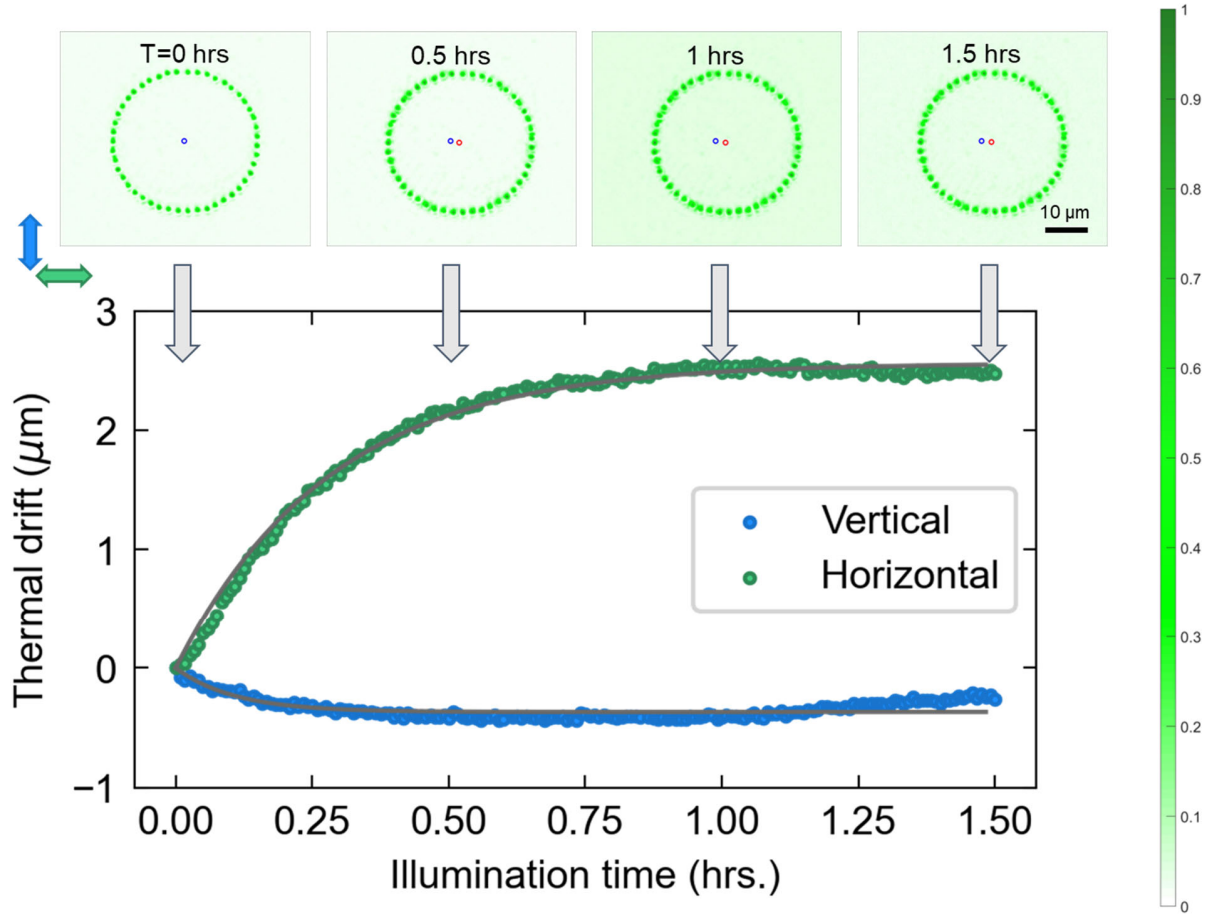

**Figure 7**: Measured thermal drifts of a polarization-switchable ring array, with optical images of the trap array at different times displayed in the top row. Blue circles indicate the center of the ring at the onset of the measurement and red circles indicate the center of the ring at the given time.

We believe that the drifts along the transverse and longitudinal directions likely originated from the thermal equilibration of one or more optical components in the characterization setup, including refractive lenses, wave plates, and the high-power laser. The drifts cannot be attributed to heating of metasurface holograms due to several reasons: (1) their small thermal capacitance corresponds to a time constant of temperature change of less than 1 minute; (2) the temperature of metasurfaces rose by less than 20°C according to our measurement using a thermal camera; (3) thermo-optical effects would introduce a nonuniform distribution of phase error (dependent on geometries of meta-units comprising a metasurface), leading to degradation of the generated holographic arrays, instead of a drift of the entire array. Once thermal equilibrium was reached, generally after ~30 minutes, the trap arrays were stable. This thermal stability is highly desirable for cold atom applications such as atomic clocks, which require an extended period of high-power illumination.

We studied the positioning accuracy of the traps by extracting the locations of individual trap sites and conducting a statistical analysis of the distance between adjacent traps. For example, the measured average spacing between adjacent traps of the ring array with designed spacing of 1.5 μm is 1.52 μm (**Fig. 3(b)**), while that of the square array with a designed lattice constant of 1.8 μm is 1.82 μm along both the x- and y-axes (**Figs. 8(c) and 8(d)**); the standard deviations of the measured lattice constant are approximately 50 nm. The deviation of the measured spacing from the design and the variation are small compared to the wavelength and can be further reduced by using higher spatial resolution during the design of the metasurfaces (spatial discretization of both the metasurface and holographic planes are 70 nm in current designs).

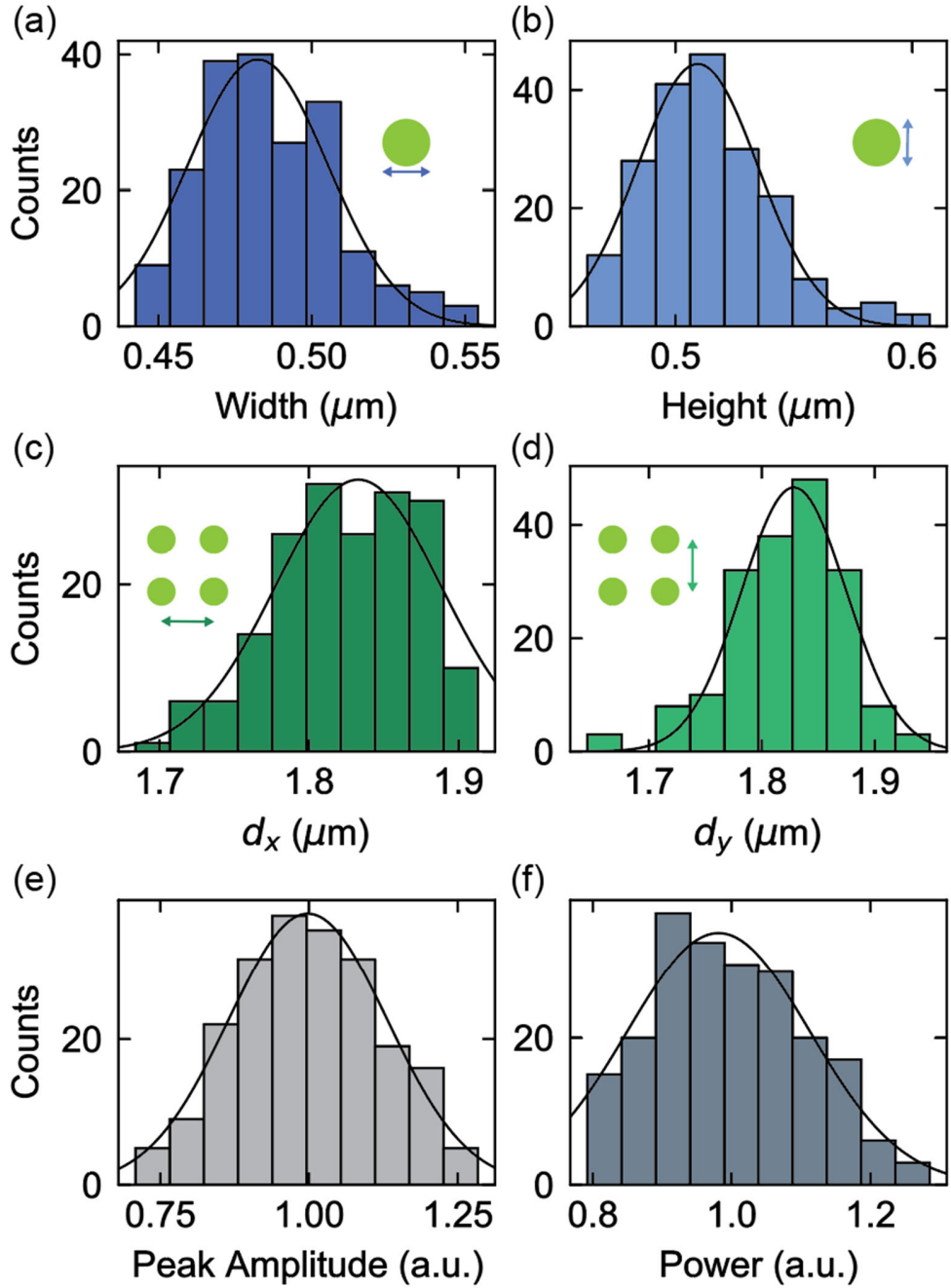

**Figure 8**: Statistical histograms of measured geometrical parameters and intensities of optical traps generated by metasurfaces. (a) and (b) are the distributions of the width and height of individual traps, respectively, of the 14×14 square lattice with a designed lattice constant of 1.8 μm. (c) and (d) are the distributions of the lattice constant of the 14×14 square lattice along the horizontal and vertical directions, respectively. (e) and (f) are the distributions of the peak intensity and integrated power of the quasi-crystal trap,

respectively. The integrated power of each trap site is evaluated by summing the pixel count over the trap.

We further characterized the size and intensity uniformity of trap arrays by extracting individual optical spots and using Gaussian fits to approximate their intensity profiles. In all traps studied, optical spots have full width at half maximum (FWHM) close to diffraction limited values. A statistical analysis was then performed to obtain the mean value and standard deviation of the sizes and peak intensity of the traps. The results for the 2D Penrose-tiling type quasi-crystal lattice are shown in **Fig. 8(e)** and **(f)** as an example. We summarize our results on all 1D and 2D traps in **Table 2**, where variation percentage is defined as the ratio between standard deviations and their corresponding mean values. In all arrays, we observed high geometrical uniformity, with the variation of spot sizes between 3% and 5%. The highly uniform size of the trap spots reduces the difference in trapping frequency, which, for example, is favorable to achieve efficient Raman side-band cooling of a single atom trapped in an optical tweezer. The spots are close to isotropic in shape: a majority of the arrays display less than 5% difference between spot sizes along the x- and y-axes. Such high isotropy allows trapped atoms to experience near-uniform confinement in the transverse directions. The intensity variation is between 12% and 16% for all the trap arrays (**Table 2**), while in simulation they display an intensity variation below 2%. We believe that this discrepancy comes from imperfect modeling of the metasurface holograms and errors in fabricated $TiO_2$ nanopillars, which can be improved in future fabrications.

| Geometries | Width variation | Height variation | Height-width difference | Peak variation |
|---|---|---|---|---|
| 1D-Dimerized | 4% | 3% | 0.9% | 12% |
| 1D-1.5 μm Ring | 5% | 5% | 0.7% | 14% |
| 1D-1.25 μm Ring | 4% | 5% | 3% | 16% |
| 2D-Kagome | 4% | 5% | 8% | 16% |
| 2D-1.8 μm Square | 5% | 5% | 2.4% | 14% |
| 2D-2 μm Square | 4.5% | 5.2% | 5% | 15% |
| 2D-Penrose | 3.2% | 3.9% | 2.2% | 12% |

**Table 2**: Statistics of measured geometry and intensity variations of the 1D and 2D trap arrays.

The variation in trap positions observed here (**Fig. 8**) is sufficiently small to enable the observation of interesting quantum-optical phenomena that have high demands on the position accuracy in atomic arrays. As an example, we studied the reflectance of a 2D atomic array (**Fig. 9**), taking into account realistic trap position variations as observed in experiments. An infinite 2D array with subwavelength spacing should perfectly reflect incoming light at a particular frequency [36,116,117]. Our theoretical study shows that a perfect finite-sized array of 14×14 atoms similarly reflects light with high efficiency (**Figs. 9(a)** and **9(b)**). We then added trap positional inaccuracy as non-correlated Gaussian noise to each trap site. We observed that while this positional inaccuracy yields reduced reflectance, the optical response still shows strong collective behavior (**Figs. 9(a)** and **9(c)**), indicating that the performance of arrays generated by metasurfaces will allow for the study of interesting quantum optics problems.

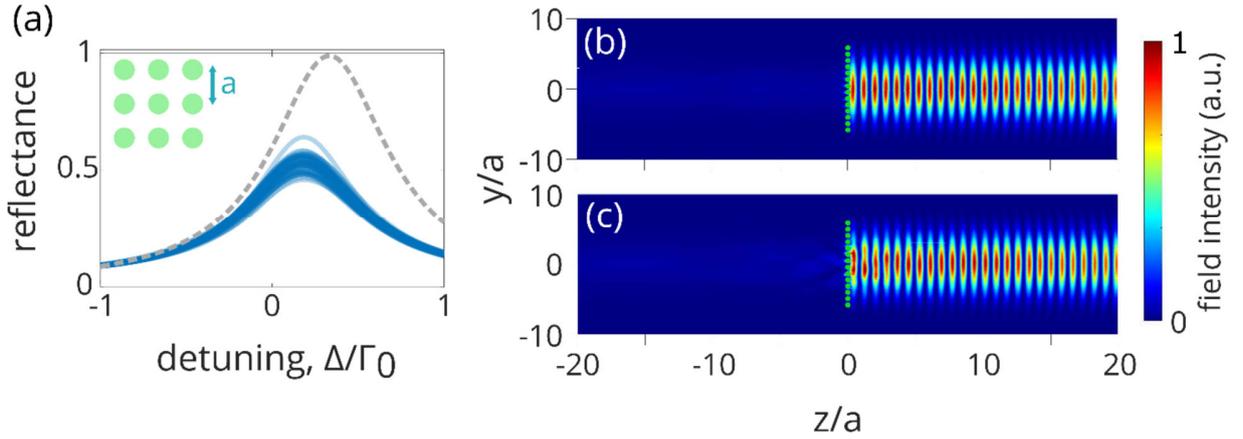

**Figure 9**: Mirror formed by a 2D atomic array with a subwavelength lattice constant. (a) Simulated reflectance of an atomic array with perfect positioning (gray dashed line) compared to 50 different realizations with positional imperfections (blue solid curves). (b,c) Spatial profiles of the electric field intensity for a perfect mirror and a realization of the imperfect case, respectively. (b,c) are taken at the detuning that yields maximum reflectance. The 2D array consists of a square lattice of 14×14 atoms with a lattice constant of 1.8 μm and positional variance of 50 nm. The lattice is driven by circularly-polarized light propagating in the z-direction. For the atoms we consider the cycling transition of $^3P_2 \rightarrow ^3D_3$ line in bosonic Strontium ($^{88}$Sr) with a resonant wavelength of 2.9 μm. The stretched state with polarization matching that of the driving field is considered to be isolated by the presence of a magnetic field, reducing the atoms to an array of circular dipoles.

The closest spacing between traps achievable with our platform is determined by the wavelength and NA of the metasurface. As a diffraction-limited system, the smallest feature achievable using metasurface holograms is constrained by the Abbe limit: $d =$

$\frac{\lambda}{2NA}$. For the metasurfaces we are currently using, the smallest spacing achievable is estimated to be ~2-2.5 times the wavelength, or 1-1.25 μm. We fabricated metasurfaces to produce ring arrays with 1-μm spacing and observed that although individual spots are discernible, the evanescent tails of adjacent spots strongly overlap, which would allow atoms to tunnel between trap sites. The array with the smallest spacing and satisfactory quality for individual spots that we demonstrated is thus the ring array with 1.25-μm spacing (**Fig. 3(c)**). We note that we did not pursue high NA for our metasurface holograms as the objective lens used to characterize the trap arrays has to withstand high optical power and has an NA of 0.6; to precisely characterize the trap arrays, the NA of the metasurfaces was chosen slightly below that of the objective lens.

Worth noting is that in a realistic cold atom experimental setup the NA of the optical system for producing trap arrays is not constrained by the metasurfaces but rather by the relay optics (e.g., a compound objective lens) used to transfer the array generated by a metasurface hologram into the vacuum chamber. Trap spots with spacing comparable to the wavelength and minimum overlap between their evanescent tails are within reach if the NA of the relay optics and that of the metasurface are increased simultaneously. Alternatively, as metasurfaces are compatible for use in vacuum, a high-NA metasurface hologram can be directly integrated into the cold atom chamber to trap atoms in a projection-optics-free manner.

We investigated the trapping frequency of optical spots generated by metasurface holograms. When atoms are positioned at the center of a focused Gaussian beam, they experience a harmonic trapping potential with trapping frequencies along the radial (transverse) and the axial (longitudinal) directions given by $\omega_r = \sqrt{4U/mw_0^2}$ and $\omega_z = \sqrt{2U/mz_R^2}$, respectively, where $w_0$ is the minimal beam waist, $z_R$ is the Rayleigh range determined by $\pi w_0^2/\lambda$, $m$ is the atomic mass, and $U$ is the trap depth, determined by the peak intensity of the beam. For non-Gaussian beam profiles, a harmonic oscillator approximation can be applied around the potential minimum and similar trapping frequencies can be extracted. In the example shown in **Fig. 10**, we simulated a kagome lattice by shining a Gaussian beam with waist of 220 μm onto a metasurface hologram of the same design as that produces the lattice in **Fig. 4(b)**. **Figure 10(c)** shows the simulated longitudinal intensity profile of one of the traps. The transverse beam waist averaged over all trap spots of the lattice as a function of the propagation distance z is shown in **Fig. 10(a)**. The minimal beam waist $w_0$ is ~0.587 μm, which can be used to calculate the radial trapping frequency $\omega_r$. The peak intensity averaged over all trap spots of the lattice along the longitudinal direction is shown in **Fig. 10(b)**. We observed that the trap beams generated by the metasurface deviate from a Gaussian beam with the same minimal beam waist (**Figs. 10(a)** and **10(b)**). We fitted a harmonic function to the ±2 μm interval about z=0 of the peak intensity profile in **Fig. 10(b)** to calculate the axial trapping frequency $\omega_z$. We found that for the kagome lattice the ratio between the radial and the axial confinement, characterized by $\omega_r/\omega_z$, is 7.3, while it is 5.0 for a Gaussian beam with the same minimal beam waist. Thus, our traps have weaker confinement along the longitudinal direction, by a factor of 1.46, compared to the Gaussian beam.

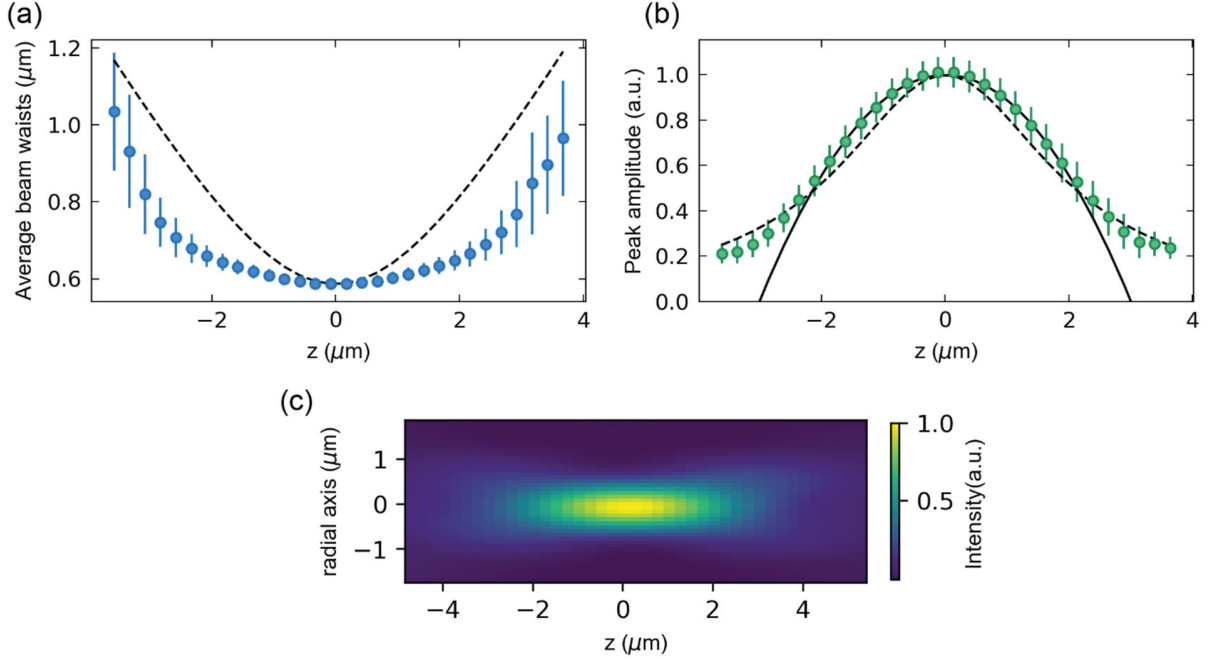

**Figure 10:** (a) Beam waist averaged over all trap spots of a simulated kagome lattice as a function of the propagation distance z. (b) Peak intensity averaged over all trap spots of the lattice a function of z. (c) Simulated longitudinal intensity distribution of one of the trap spots. The error bars in (a) and (b) account for the variation of the trap spots across the lattice and fitting errors. The black dashed curves in (a) and (b) are the profiles of an ideal Gaussian beam with the same minimal beam waist of 0.587 μm. The black solid line in (b) is a fit to the data (the ±2 μm interval about z=0) by a harmonic function $U_0(1 - \frac{1}{2}k(z-z_0)^2)$.

We summarize the features of our metasurface holographic traps and compare them with traps created by AOD and SLM setups in **Table 3**. The distinct advantages of metasurface holograms include their ability to produce arbitrary array geometries at arbitrary wavelengths with close spacing between traps, compact footprint, high optical efficiency, and excellent power handling capability.

| Characteristics | AOD | SLM | Metasurface |
| --- | --- | --- | --- |
| Type | Active | Active | Passive |
| External power | Yes | Yes | No |
| Relay optics | Yes | Yes | No |
| Wavelength range | Specified by model | Specified by model | Arbitrary |
| Power handling | 10 W/cm$^2$ | 200 W/cm$^2$ | >1200 W/cm$^2$ |
| Device footprint | Tens of centimeters | Tens of centimeters | Millimeter |
| Spacing between traps | Loose (~ 3 μm at λ=520 nm) | Dense (e.g., ~1.25 μm at λ=520 nm) * | Dense (e.g., ~1.25 μm at λ=520 nm) |
| Trap geometry | 2D simple geometry | Arbitrary pattern in any dimension | Arbitrary pattern in any dimension |
| Power efficiency | ~50% | ~40% | ~60% (up to 80% in simulation) |
| Peak variation | ~3% | ~3% | So far: ~12% |

**Table 3**: Comparison between conventional platforms to generate optical trap arrays based on AODs and SLMs, and our platform based on metasurface holograms. *: with external focusing optics.

As mentioned in **Section 1**, AODs, SLMs, and DMDs have been conventionally used to produce optical traps for atomic arrays. AOD-based methods, while capable of generating patterns with high uniformity (via driving with multiple RF tones), are only able to realize relatively simple geometries such as periodic patterns. SLMs and DMDs are capable of generating trap arrays with arbitrary geometries in a dynamically tunable fashion, which makes it possible to iteratively improve upon the quality of the generated traps using real-time feedback. However, the approaches based on SLMs and DMDs suffer from relatively low spatial resolution, limited power handling capability, presence of a zeroth-order diffractive background, and large footprint. In particular, the large spatial pitch (pixels with a linear dimension of 5-10 μm) of SLMs and DMDs limits their ability to produce high spatial frequency Fourier components (i.e., light deflected at large angles) needed to form high-quality traps. This results in low NA and large optical spots, hampering their ability to generate traps with close spacing. To address this issue, a high-NA objective lens has to be used to demagnify the optical pattern generated by SLMs and DMDs, increasing the dimension and complexity of the overall experimental setup.

While SLMs can achieve close spacing between trap sites with the help of a high-NA objective, their large pixel sizes limit the total number of pixels to about one million, given a certain SLM footprint and complexity of the control electronics. This limitation precludes the generation of large-scale, high-quality arrays with thousands of trap sites desired for quantum simulation and optical clock applications. In comparison, there is no fundamental limit on the number of meta-units or nanopillars constituting a metasurface (typical metasurfaces already contain from 1 to 10 million meta-units and the number can be readily scaled up), making it possible to generate large-scale arrays with tens of thousands of trap sites. Another consideration is that generating exceedingly large arrays while maintaining substantial optical intensity at each trap site requires high power-handling capability. In both regards, the scalability and high thermal stability of metasurfaces make them highly suitable for such tasks.

From a practical standpoint, the requirement of external power supplies and cooling mechanisms for SLMs and AODs makes them ill-suited for the growing need of integration and miniaturization necessary for wide deployment of portable atomic systems such as atomic clocks. Also, these complex optical devices are more sensitive to vibrations, making them less ideal for real-world applications. Finally, SLMs and AODs are not as affordable or accessible as metasurfaces, which can be readily fabricated on a large scale using well-established CMOS-compatible nanofabrication techniques in modern foundries.

6.  **Future perspectives**

One of the limiting factors in the wide deployment of portable cold-atom-based devices and systems to address real-world applications has been their footprints. Conventional AOD and SLM setups typically require large and complex optical circuits with extra power supply and cooling mechanisms, making them bulky and susceptible to environmental interference. Our approach based on metasurfaces has the potential to evolve into an ultra-compact solution where high-NA metasurface holograms are directly incorporated into the vacuum chamber to generate optical trap arrays without the assistance of additional optical components. Currently, the size of our metasurface holograms (with a linear dimension of ~500 $\mu$m) is limited by the slow writing speed of electron beam lithography; thus, given a target NA of ~0.45, the working distance of the holograms (i.e., distance between a metasurface and the holographic pattern it generates) is ~500 $\mu$m. Recent advances in planar fabrication techniques, especially deep-UV photolithography, have enabled the creation of wafer-sized (up to 10 cm in diameter) metasurfaces operating in the near-infrared [118] and visible [119,120] regimes. These advances make it possible to realize high-NA metasurface holograms with long working distances that can be operated outside of the vacuum chamber to generate arrays to trap cold atoms within the chamber. For example, a metasurface hologram with a linear dimension of 2 cm and NA of 0.75 can generate a trap array, with spots spaced by one wavelength and non-overlapping evanescent tails, ~1 cm away from the hologram, on the opposite side of the wall of a vacuum chamber. This approach may be preferred in experiments because it alleviates a number of complexities, including reaching ultra-high vacuum with metasurface chips and their holders in the chamber, difficulty and time cost of exchanging the chips, management of laser heating of the chips in the vacuum, and deposition of atoms on the chips over time.

While we have shown that interesting physics can be studied within the measured performance of the metasurface generated arrays, an important goal is the further improvement of position and intensity uniformity. There are a few improvements that one can adopt to mitigate current variations, including better thin-film growth, modeling, and fabrication techniques. In particular, our current metasurface design protocol is based on the locally periodic approximation, which assumes that individual meta-units would behave as if they were placed in an infinite periodic array of meta-units; in actual metasurfaces, however, a meta-unit could be surrounded by nanopillars with drastically different shapes, sizes, and phase responses. As a result, near-field coupling between neighboring nanopillars could lead to deviation from their simulated optical responses [121]. We are developing a coupled-mode-theory-based inverse design methodology that takes into account explicitly such coupling in the design and simulation of metasurfaces. In particular, the nanopillars are modeled as truncated waveguides that couple by modal overlap with their neighbors and their amplitude and phase responses can thus be calculated in an accurate and time-efficient manner; a stochastic-gradient-descent-based algorithm is then employed to find optical metasurface designs that minimize a global loss function. With this approach, the performance of metasurfaces can be accurately modeled and efficiently optimized so that we expect that intensity uniformity of the optical traps obtained in experiments will closely follow calculations.

If power handling or portability of a system is not a concern, it is also conceivable that a combination of a metasurface hologram and a DMD or SLM (with a low pixel count and thus economic one) could enable the generation of large arrays with improved intensity uniformity compared to using the metasurface alone. The design of a metasurface hologram is based on a certain input Gaussian beam illumination; an actual incident Gaussian beam is likely an imperfect one with its beam size, and phase and intensity profiles not precisely matching with the assumption, or even with intensity ripples not considered in modeling. The DMD or SLM can be used to provide low-resolution reshaping of the Gaussian beam before its interaction with the metasurface.

The metasurface holograms demonstrated in this work have been designed to realize a single function at a particular wavelength. With more complex meta-unit archetypes and more advanced metasurface design concepts, it is possible to incorporate multiple functionalities at different wavelengths into a single metasurface. As an example, by properly choosing the cross-sectional shape of nanopillars, we are able to control the effective modal index of the nanopillars as a function of wavelength and polarization state (i.e., dispersion engineering of meta-units) [10,11]; this capability will allow us to create a metasurface that projects an optical trap at one wavelength (or at one polarization state) and images fluorescence signals at another wavelength (or at the orthogonal polarization state) [122]. As another example, by introducing several spatially distributed, symmetry-breaking perturbations into a 2D photonic crystal lattice, we can create a novel wavelength-selective metasurface supporting multiple quasi-bound states in the continuum that is able to mold the optical wavefront into distinct shapes at selected wavelengths, while leaving the optical wavefront at other wavelengths unchanged [96,97]. This capability will allow us to realize a "color-multiplexing" scheme for quantum optics experiments, where multiple laser beams can be independently shaped, for cooling, trapping, and monitoring cold atoms, yet share the same optical port.

## 7. Conclusion

In summary, we have demonstrated a variety of optical trap arrays at $\lambda$=520 nm (for cold Strontium atoms) via metasurface holograms and characterized their figures of merit, including the homogeneity and positioning accuracy of the traps, and power handling capability of the metasurfaces. We have experimentally shown that the generated holographic traps possess high positioning accuracy, size uniformity, optical efficiency, and thermal stability. Importantly, our metasurfaces can directly generate high-NA arrays with desired geometries without help from other optical elements, such as high-NA objectives commonly used in conjunction with SLMs. These features make them highly desirable for applications that demand high-quality atomic arrays with dense spacing between hundreds and thousands of trap sites in a compact system. The high-performance and compactness of metasurface holograms and the possibility of leveraging CMOS foundries for low-cost, high-throughput, and high-yield fabrication of such metasurfaces pave way for economic, portable deployment of metasurface-generated arrays in real-world application scenarios, a prospect not possible with existing methods.

**Acknowledgements**

This work was supported by the National Science Foundation (grant no. QII-TAQS-1936359, no. CA-2040702 and no. ECCS-2004685) and the Air Force Office of Scientific

Research (grant no. FA9550-16-1-0322). Device fabrication was carried out at the Columbia Nano Initiative cleanroom, at the Advanced Science Research Center NanoFabrication Facility at the Graduate Center of the City University of New York, and at the Center for Functional Nanomaterials, Brookhaven National Laboratory, which is supported by the US Department of Energy, Office of Basic Energy Sciences (contract no. DESC0012704). W. Yuan thanks the Croucher Foundation for its funding support.